\documentclass[twocolumn,secnumarabic,amssymb, nobibnotes, aps, pre,superscriptaddress]{revtex4}
\usepackage{amsmath,graphicx,float}

\begin{document}
\bibliographystyle{unsrt}

\title{The Fourier heat conduction as a strong kinetic effect}

\author{Hanqing Zhao}
\affiliation{Department of Modern Physics, University of Science and Technology of China, Hefei 230026, China }
\affiliation{School of Physical Science and Technology, and Key Laboratory for Magnetism and Magnetic Materials of MOE, Lanzhou University, Lanzhou, Gansu 730000, China}
\author{Wen-ge Wang}
\affiliation{Department of Modern Physics, University of Science and Technology of China, Hefei 230026, China }

%\date{january 18, 2017}%
\begin{abstract}
For an one-dimensional (1D) momentum conserving system, intensive studies have shown that generally its heat current autocorrelation function (HCAF) tends to decay in a power-law manner and results in the breakdown of the Fourier heat conduction law in the thermodynamic limit. This has been recognized to be a dominant hydrodynamic effect. Here we show that, instead, the kinetic effect can be dominant in some cases and leads to the Fourier law. Usually the HCAF undergoes a fast decaying kinetic stage followed by a long, slowly decaying hydrodynamic tail. In a finite range of the system size, we find that whether the system follows the Fourier law depends on whether the kinetic stage dominates. Our study is illustrated by the 1D diatomic gas model, with which the HCAF is derived analytically and verified numerically by molecular dynamics simulations.
\end{abstract}
\maketitle

Driven by applications of nanomaterials, heat conduction properties of low-dimension materials have been a focus topic in the past three decades~\cite{graphene, nanotube, lepri, dhar, linianbei, lebow, wang}. Based on numerous theoretical studies, it is concluded that in general, the thermal conductivity of a low-dimensional momentum conserving system has a system-size dependent abnormality in the thermodynamic limit~\cite{spohn,beijeren,chen1}.  This abnormality is attributed to the hydrodynamic effect that induces the slow power-law decay of the heat current autocorrelation function (HCAF)~\cite{beijeren, spohn}. However, counterexamples, i.e., momentum conserving systems but yet having a size-independent thermal conductivity, have been found~\cite{chen1, chen2, zhong, savin}. These counterexamples usually have asymmetric interparticle interactions, including the Toda-like models~\cite{zhong}, Lennard-Jones model~\cite{chen2, savin, chen1}, the Fermi-Pasta-Ulam-$\alpha$-$\beta$ model~\cite{saito}, the diatomic gas model~\cite{chendiatom, chendiatom2}, the diatomic Toda model~\cite{chendiatom2}, and so on. More importantly, so far direct experimental measurements have not provided solid evidence yet to support the predicted abnormality in real low-dimensional materials of finite sizes available.

In one dimensional (1D) case, the heat conductivity, denoted by $\kappa$, is related to the HCAF, denoted by $C(t)$, by the Green-Kubo formula
\begin{equation}
\kappa=\lim_{t_c\to\infty}\lim_{L\to\infty}\frac{1}{k_BT^2L}\int_0^{t_c} C(t)\,dt.
\end{equation}
Here $k_B$ is the Boltzmann constant, $L$ and $T$ are, respectively, the size and the temperature of the system,
and $C(t)\equiv \langle J(0)J(t)\rangle$, with $J(t)$ being the total heat current at time $t$ and $\langle \cdot \rangle$ representing the equilibrium ensemble average. For a finite system, Lepri {\it et al}.~\cite{lepri} suggest to drop the limits and truncate the integral at $t_c=L/c_s$ ($c_s$ is the sound speed) to calculate the heat conductivity~\cite{lepri,chen1}. It leads to $\kappa\sim L^{1-\alpha}$ in the thermodynamical limit
given that $C(t)$ tends to decay as $\sim t^{-\alpha}$ as $L\to \infty$.

In trying to understand the aforementioned counterexamples and the existing experimental results, Chen {\it et al}. conjectured that the asymmetric interactions may practically lead to a size-independent thermal conductivity in a certain finite system size range~\cite{chen1}, because the hydrodynamic approach may not apply to systems of asymmetric interactions in a transient, but may be long time period. In this transient period, the HCAF may decay faster, but its contribution to the thermal conductivity can dominate until that contributed by the hydrodynamic power-law tail becomes comparable after a sufficient long time. Therefore, though the predicted abnormality can be the case in the thermodynamics limit, normal heat conduction following the Fourier law can still be expected in a finite system size range. This would have significant practical implications because any real materials are in fact finite.

It is thus important to establish a complete theory based on which the kinetic effect can be evaluated and taken into account as well. This is our motivation and in this work, we will focus on the 1D diatomic gas~\cite{casatidiatom}, a paradigmatic, momentum conserving fluid model. The hard-core elastic collision occurring when two neighboring particles meet can be considered as an effective asymmetric interaction. It is worth noting that there has been a long-term argumentation towards the heat conduction property of this model. In 2001, Garrido {\it et al}. presented the numerical evidence to show that this model has a convergent thermal conductivity in the thermodynamic limit~\cite{diatom1}. This result was questioned by many other authors~\cite{dhar2, lihaibin, yang} because of the clear power-law decaying tail in the HCAF. Nevertheless, a recent numerical study showed that interestingly, when the two types of particles in the system have close masses, the heat conductivity does not depend on the system size in a certain system size range~\cite{chendiatom2}. Moreover, it is observed that the HCAF shows an exponential-like decay in a transient stage, and the time this transient stage lasts increases rapidly as the mass ratio tends to 1. Therefore, the mass ratio is a key parameter for the heat conduction property of this model. In view of the subtlety of this issue and the limitation of the numerical simulations, an analytical study is particularly desired. In the following we will show that the idea of the conventional kinetic approach can be borrowed for our aim here.

First of all, suppose that our model consists of $N$ particles with alternative masses $\mu_1$ and $\mu_2$ queueing on a line. We assume that $\mu_1>\mu_2$ and define $r=\mu_1/\mu_2$ as the mass ratio. Let $m_i$ and $v_i$ be the mass and velocity of the $i$th particle; after a collision, the velocities of the two neighboring particles, say the $i$th and the ($i$+1)th, change into
\begin{align}
    v_i(t+1)&=&\frac{m_i-m_{i+1}}{m_i+m_{i+1}}v_i(t)+\frac{2m_{i+1}}{m_i+m_{i+1}}v_{i+1}(t),\nonumber\\
    v_{i+1}(t+1)&=&\frac{m_{i+1}-m_i}{m_i+m_{i+1}}v_{i+1}(t)+\frac{2m_i}{m_i+m_{i+1}}v_{i}(t),
\end{align}
where the time $t$ is measured as the number of collisions. Note that this dynamics keeps the total momentum and energy of the system.

In the kinetic theory, for characterizing the Brownian motion, one traces a tagged particle and studies the decay of its velocity autocorrelation function. In our  model a particle is always bounded by its two neighbors. Hence, instead tracing a tag particle, we record the energy the tagged particle carries initially, which we term as the `tagged energy', and investigate how it spreads over the system. Note that during a collision, the energy carried by a particle will separate into two parts; one part remains on itself, while another part transfers to the other particle. If the two particles have close masses, then the transferred part will hold a large proportion. For the sake of convenience, in the following we term this transferred part of energy as the dominant energy since it dominates the decay behavior of the HCAF before the hydrodynamic process takes over. Similarly, we term the particle that carries the dominant energy as the dominant carrier. (At a given time there is only one dominant carrier.) Then the tagged energy can be traced by following the ensuing dominant energy sequence and  in turn by tracing the dominant carriers. This is the key technique we adopt for our analytical treatment, which can be seen as an extension of the conventional kinetic approach.

As an example, let us take the $i$th particle as the tagged particle and assume its first collision happens with the ($i$+1)th particle. According to Eq.~(2), after the collision its initial velocity $v_i(0)$ separates into two parts, the remaining part $\frac{m_i-m_{i+1}}{m_i+m_{i+1}}v_i(0)$ and the transferred part $\frac{2m_i}{m_i+m_{i+1}}v_i(0)$. For $m_1\approx m_2$, the remaining part will be much smaller than the transferred, the $(i+1)$th particle thus carries the dominant energy and becomes the dominant carrier. As the velocity component $\frac{2m_i}{m_i+m_{i+1}}v_i(0)$ in $v_{i+1}(1)$ comes from $v_i(0)$, it correlates with $v_i(0)$. Similarly, when the next collision happens between the $(i+1)$th particle and one of its neighbors (no matter the $i$th or the ($i$+2)th as they have the same masses), the transferred  velocity that contains $v_i(0)$ is $\frac{4m_{i} m_{i+1}}{(m_i+m_{i+1})^2}v_i(0)=\frac{4r}{(1+r)^2}v_i(0)$. It is thus straightforward that after $2P$ collisions, the portion of $v_i(0)$ that transferred to the dominant carrier is
\begin{equation}
v_i(0)[\frac{4m_{i} m_{i+1}}{(m_i+m_{i+1})^2}]^P=v_i(0)[\frac{4r}{(1+r)^2}]^P.
\end{equation}

Now let us consider the HCAF. The total energy current is defined as $J(t)\equiv \sum _{q}^N j_q(t)$, where $j_q(t)\equiv \frac{1}{2}m_q v_q^3(t)$  is the local current on the $q$th particle. Still taking the $i$th particle as the tagged particle, we have~\cite{lepri,zhong}
\begin{equation}
\langle J(t)J(0)\rangle=N\sum _{q}^N\langle j_q(t)j_i(0) \rangle,
\end{equation}
Suppose that at time $t=2P$ the the dominate carrier is the $k$th particle and $t_F$ is the average time for the dominant energy transferring from one dominant carrier to the next; For $r\to 1$, we have $j_q(t)j_i(0)\neq0$ for $q=k$ and other $j_q(t)j_i(0)$ terms are negligible. This gives that
\begin{equation}
C(t)=N[\frac {64r^3}{(r+1)^6}]^{\frac{t}{2t_F}}\langle j_i(0)j_k(0)\rangle,
\end{equation}
which can be rewritten as
\begin{equation}
\langle C(t)\rangle=\langle C(0)\rangle e^{-\frac{t}{\tau}}
\end{equation}
with
\begin{equation}
\tau=-2t_F[\ln(\frac{64r^3}{(r+1)^6})]^{-1}.
\end{equation}

%\begin{figure*}
%\centering
%\includegraphics[width=18.7cm]{fig1.eps}
%\caption{Comparisons between the predicted and simulated results. (a)The HCAF at %different mass ratio. Soloid lines are simulation results and dash lines are theoretical %prediction. (b) is the thermal conductivity. The solid-dot lines are integrals of simulated %HCAF by Green-Kubo formula, and the horizen dash lines are the $\kappa_{k}$. (c) shows the %turning point $t_c$. The green crisscrosses are simulation results and the bluecrisscorsses %are theoretical prediction. }
%\end{figure*}

Equations (6) and (7) are our main result, which indicate that the HCAF decays exponentially in the kinetic stage. The physics picture behind Eq.~(6) is indeed similar to that of the Brownian motion. In our case, the tagged particle is bounded, but its energy is dispersed due to the interactions with and among the surrounding particles, resulting in an exponentially decaying energy current autocorrelation function. In our model, the dominant carrier changes from one to another. During this process, the tagged energy keeps losing, so that the HCAF, dominated by remained energy from dominant carriers, decays exponentially. Therefore, our treatment is the same in spirit as the conventional kinetic approach. The parameter to be determined is $t_F$. As it is generally accepted that the energy is transported by the sound modes~\cite{hansen}, it is reasonable to assume that the energy is transferred at the sound speed as well, and therefore we have $t_F =\frac{a}{c_s}$, where $a$ is the average distance between two neighboring particles. (Throughout this work we set $a=1$ so that $N=L$.)

By substituting (6) into the Green-Kubo formula (1), we can obtain the thermal conductivity due to the kinetic effect exclusively:
\begin{equation}
\kappa_{k}=\frac{\tau}{k_B T^2}C(0).
\end{equation}
When the hydrodynamic contribution is negligible in a finite system, we have $\kappa\approx \kappa_k$.  In fact, as the dominant energy keeps losing [see Eq.~(3)], the energy transferred to other particles can not be neglected after a sufficient long time. This part of energy evolves following the hydrodynamics and can be captured by the hydrodynamics approaches~\cite{beijeren}. The decaying behavior of the HCAF induced by the hydrodynamics process has been worked out~\cite{spohn,beijeren,spohn2}, which reads $C_H(t)=ct^{-\frac{2}{3}}$ in the thermodynamic limit.  The parameter $c$ is the amplitude of the power-law tail. Roughly, we can identify the time, denoted by $t_1$, that separates the kinetic and the hydrodynamic stage by the condition
\begin{equation}
C(0) e^{-\frac{t_1}{\tau}}=ct_1^{-\frac{2}{3}}.
\end{equation}
% After $t_1$ the HCAF is affected more by other particles than by the dominant carriers. If the kinetic region is big enough, which takes place at $r\rightarrow 1$, the HCAF will decay for several orders in amplitude at the kinetics dominant region. In this case, though the hydrodynamic contribution exceeds the kinetics contribution after the turning point $t_1$ , it is negligible comparing to $\kappa_k$ till to the second turning point $t_2$ after which the hydrodynamic contribution becomes dominant.  $t_2$ can be estimated by
For $t<t_1$, the kinetic effect dominates. Because as $r\to 1$, both $\tau$ and $t_1$ increase [see Eq.~(7) and (13)], the kinetic region can last so long to allow the HCAF to decay for orders in the amplitude (see Fig.~1 for $r=1.1$ as an example). On the other hand, for $t>t_1$, the hydrodynamic effect begins to take over, but its contribution to the heat conductivity will not be comparable before another time scale, denoted by $t_2$, that can be estimated by
\begin{equation}
\kappa_H = c\int_{t_1}^{t_2}t^{\frac{2}{3}}dt.
\end{equation}
Namely, for $t>t_2$, we have $\kappa_H>\kappa_k$. The time scale $t_1$ and $t_2$ thus suggest two characteristic system sizes $L_1=c_s t_1$ and $L_2=c_s t_2$, for $L_1<L<L_2$ we can expect that the heat conductivity is in effect independent of the system size. This is consistent with the previous numerical study~\cite{chendiatom2} (see also Fig.~2 for $r=1.1$ as an example).

%Within $t_1$ and $t_2$, $\kappa$ should appear as a time-independent stage approximately. With $r\to1$,  the stage can extend to a sufficient big scale. According to the relation $t_{c}=L/c_s$, this fact implies that the thermal conduct of such a system should keep as a constant till to a sufficient large size.

    \begin{figure}[!t]
    \centering
    \includegraphics[width=8.9cm]{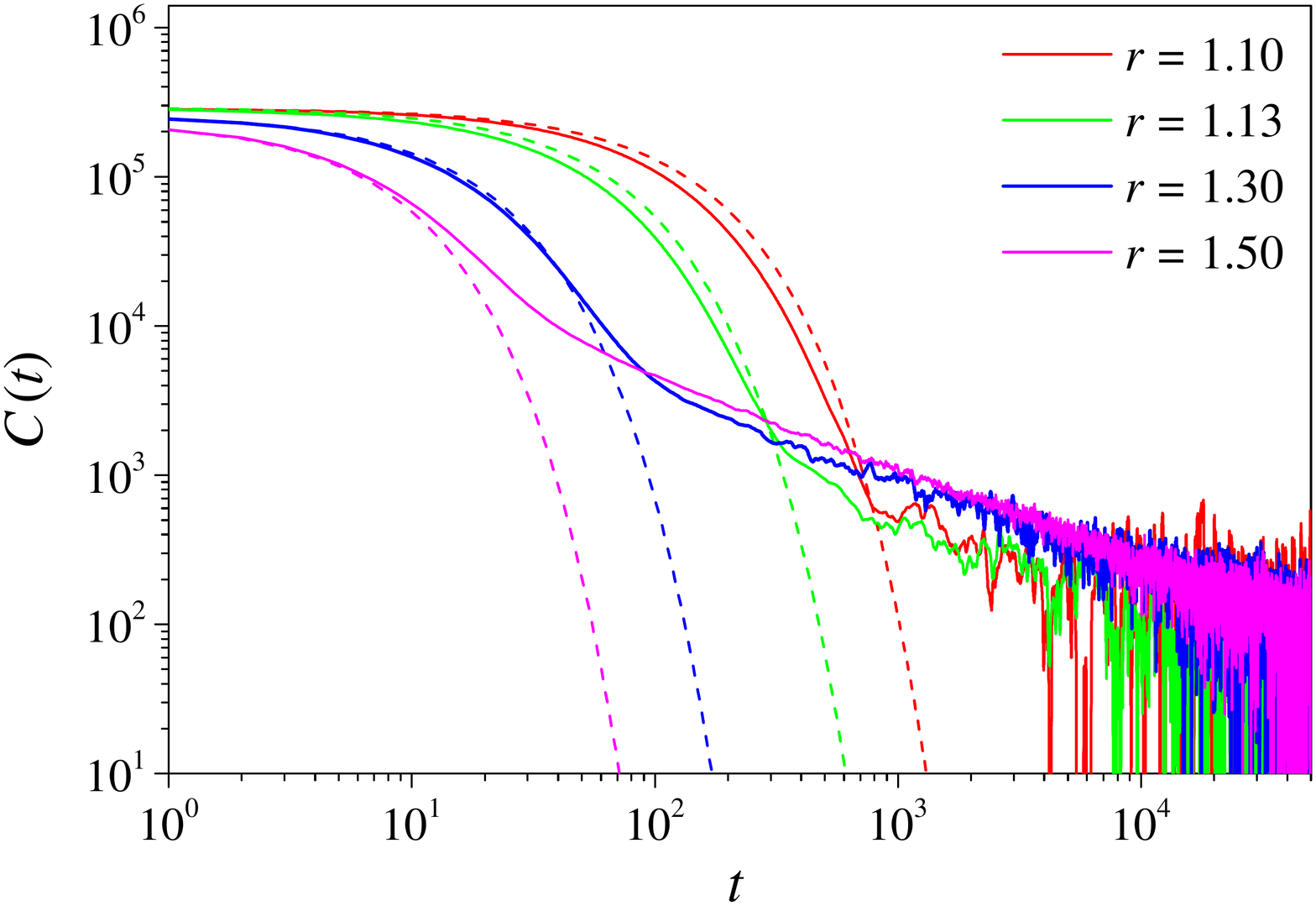}
    \caption{The heat current autocorrelation function of the 1D diatomic gas model at various mass ratios. The solid lines are for simulation results and the dashed lines of the same color are for the corresponding analytical predictions [Eq.~(6)] based on our extended kinetic theory. In simulations, the system size is set to be $N=50000$. Here and in all other figures, $k_B=1$, $T=1$ and $\mu_2=1$.}
    \end{figure}

    \begin{figure}[!t]
    \centering
    \includegraphics[width=8.9cm]{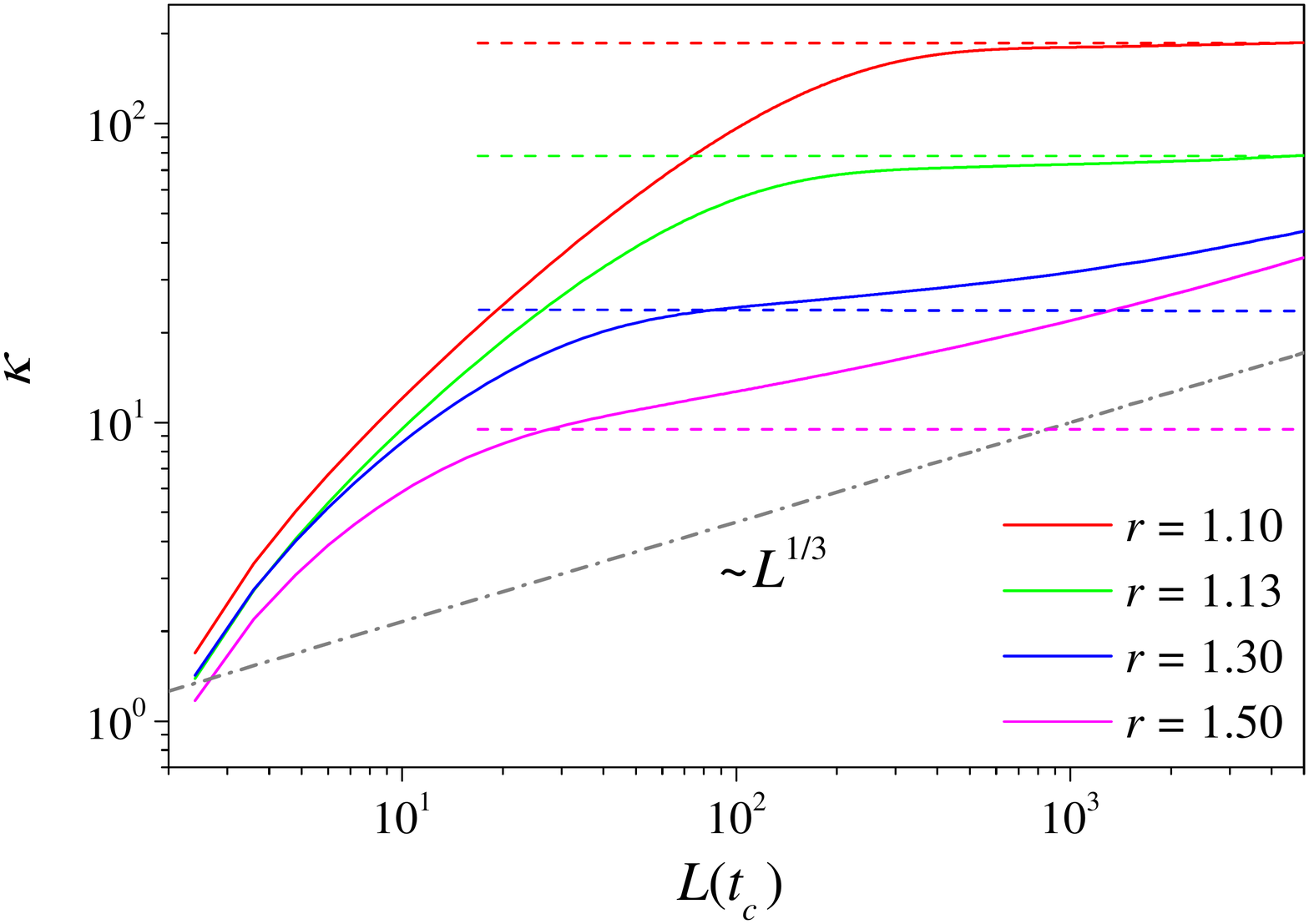}
    \caption{The heat conductivity of the 1D diatomic gas model as a function of the system size at various mass ratios. The solid lines are for the results based on the Green-Kubo formula by integrating the heat current autocorrelation function numerically obtained. The horizontal dashed lines of the same color are for the corresponding result $\kappa_{k}$ [Eq.~(8)] due to the pure kinetic effect. }
    \end{figure}

Our method can be adopted to study other 1D gas models. An immediate application is to the model with periodically repeating unit of one heavy particle of mass $\mu_1$ and $Z-1$ light particles of mass $\mu_2$. In this model,  when two light particles collide, they simply exchange their velocities. Hence the dominant energy is transferred along the $Z-1$ light particles ballistically without any decay till the collision with a heavy particle occurs. As a result, the time $t_F$ should be $Z-1$ times of that in the alternative diatomic gas, i.e., $t_F=\frac{(Z-1)a}{c_s}$. Hence Eq.~(6) and (7) do not change except that the value of $t_F$ in (7) should be replaced. 
%Hence Eq.~(6) keeps unchanged while Eq.~(7) should be replaced by
%\begin{equation}
%\tau=-(Z-1)t_F[\ln(\frac{64r^3}{(r+1)^6})]^{-1}.
%\end{equation}

Our method can be applied to models with random masses as well. Let us first consider the random diatomic gas model, where a particle has a probability of 1/2 to adjoin the same type of particles. The isotactic clusters formed have an average length of $\langle b\rangle =\sum_{i=0}^{+\infty}\frac{1}{2}^ia=2a$, which implies that on average $t_F=\frac{2a}{c_s}$. So again we only need replace the value of $t_F$ in Eq.~(7). For the more general model where all the particles have random but close masses, on average after a collision the HCAF is
\begin{equation}
\langle J(t_F)J(0) \rangle =8\langle \frac {m_{i}^2m_{i+1}}{(m_{i}+m_{i+1})^3}\rangle C(0),
\end{equation}
where $\langle \frac {m_{i}^2m_{i+1}}{(m_{i}+m_{i+1})^3}\rangle$ is determined by the mass distribution. After time $t$ the dominant carrier experienced $P$ collision, and $t=P t_F$. The HCAF in the transient kinetic process also follows  Eq.~(6) and (7) with
\begin{equation}
\tau=-t_F[\ln(\frac{m_{i}^2m_{i+1}}{(m_{i}+m_{i+1})^3})]^{-1}.
\end{equation}

    \begin{figure}[!t]
    \centering
    \includegraphics[width=8.9cm]{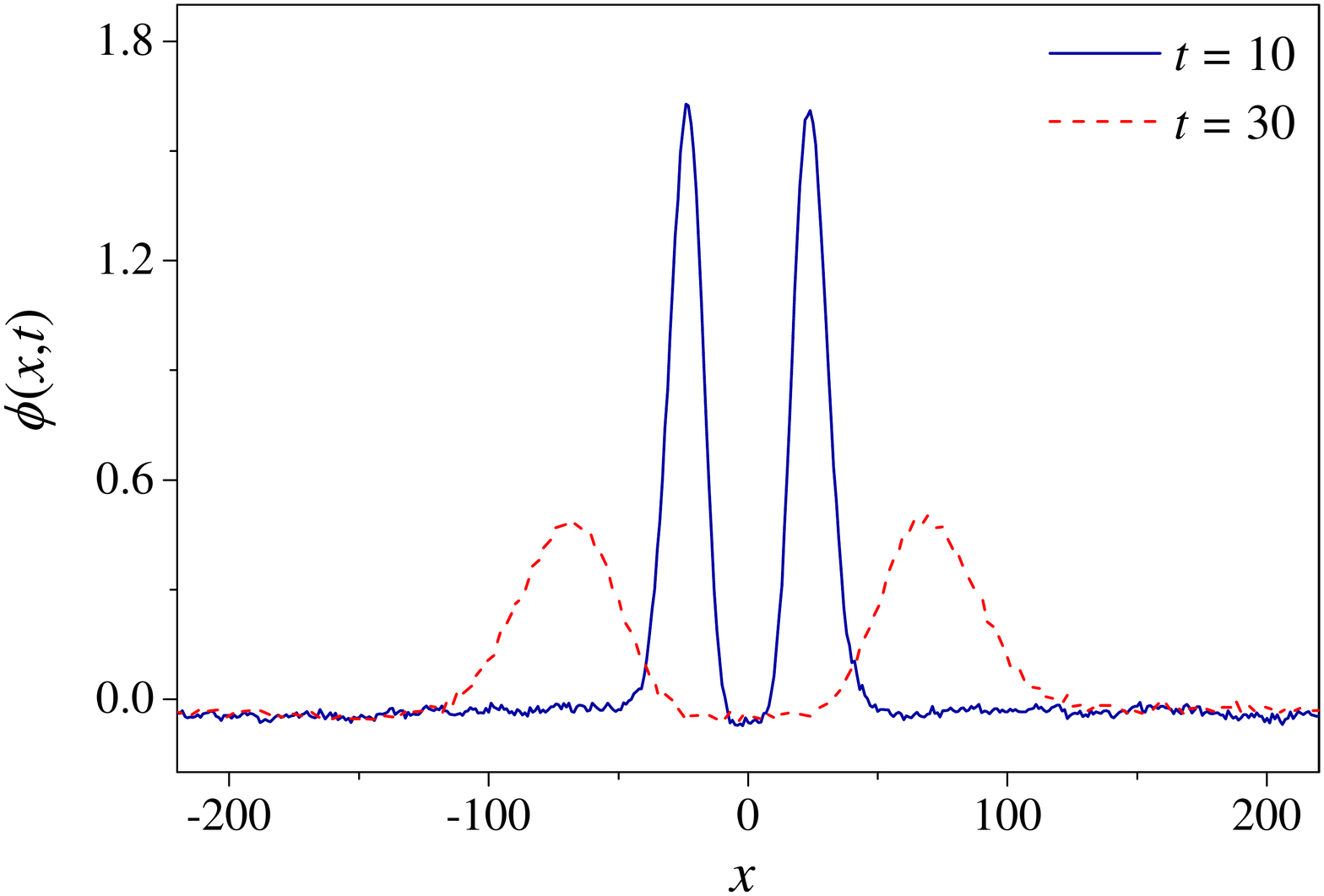}
    \caption{The numerical simulation results of the spatio-temporal correlation functions of heat current density fluctuations of the 1D diatomic model with $r=1.3$ at $t=10$ (blue line) and $t=30$ (red line) respectively. By tracing the sound mode peaks the sound speed is measured numerically.  }
    \end{figure}

Now we put our analytical results into numerical check with the 1D diatomic gas model. As how to calculate analytically the sound speed of this model is still an open problem, we evaluate the sound speed by the aid of numerical simulations as well~\cite{saito, soundspeed, zhao}. (Note that recently the sound speed for nonlinear 1D lattices with analytical interactions has been derived analytically~\cite{spohn}. It, however, does not apply to the 1D gas models.) In doing so, we measure the sound speed by tracing the motion of the sound mode peaks in the spatio-temporal correlation function of the heat current density fluctuations~\cite{zhao} (see. Fig.~3). For $1.1\le r\le 1.5$, we find that $c_s\approx 2.3$. For the HCAF, in our simulations a system consisting of 50000 particles with periodic boundary condition is considered. The numerical results and the analytical results are compared in Fig.~1 for various mass ratios. We can see that the predicted HCAF based on the kinetic effect agrees with the simulation result very well in the transient time region $t<t_1$. In Fig.~2 we plot the corresponding thermal conductivity calculated by the Green-Kubo formula as a function of the system size, where, for a given system size $L$, $\kappa$ is measured by integrating $C(t)$ up to the truncated time $t_c=L/c_s$. It shows that the kinetic approach does allow us to predict the system-size independent range of $\kappa$, which increases as $r\to 1$. For larger $r$ this range disappears though a crossover can be identified. The prediction fails for larger $r$ since the hydrodynamic effect begins to play a  significant role before the HCAF decays to a sufficient small value. For example, for $r=1.3$, the hydrodynamic contribution becomes dominant rapidly and the power-law divergence of $\kappa\sim L^\frac{1}{3}$ can be recognized for $t>10^3$(see. Fig. 2).
%In Fig.~2, a snapshot of the spatio-temporal correlation function at $t=50$ is plotted. By tracing the peak one can obtain the sound velocity. In the case with unity temperature, we find that $c_s\sim2.3$ at different mass ratios[].

Next, we check the time scale $t_1$. To evaluate it by Eq.~(9), we need to know the parameter $c$ in principle. However, the hydrodynamic approach has not solved it yet. (Note that the analytical approach developed by Spohn~\cite{spohn} applies only to the 1D lattices with analytical interactions.) Fortunately, we notice that $t_1$ does not depend on $c$ sensitively. In fact, Eq.~(9) can be rewritten as $\frac{t_1}{\tau}+\ln (\frac{c}{\langle C(0)\rangle})=\frac{2}{3}\ln(t_1))$. As for $r\to1$, $t_1$ is large, while the term $\ln(\frac{c}{\langle C(0)\rangle})$ is negligible since $\frac{c}{\langle C(0)\rangle }\sim10^{-1}$~\cite{spohn2},  we therefore can solve the following equation
\begin{equation}
\frac{t_1}{\ln(t_1)}=\frac{2}{3}\tau
\end{equation}
to estimate $t_1$. The result is shown in Fig.~4. One can see that it agrees very well with that obtained by simulations. (In simulations, $t_1$ is identified to be the convex-concave transition point of the HCAF.) From Fig.~4, it is seen that for $r>1.5$, $t_1$ drops to the order of 1, suggesting that the hydrodynamic region covers almost the entire time region. This is also consistent with previous numerical studies that claimed the abnormal heat conduction with $r>1.5$.

    \begin{figure}[!t]
    \centering
    \includegraphics[width=8.9cm]{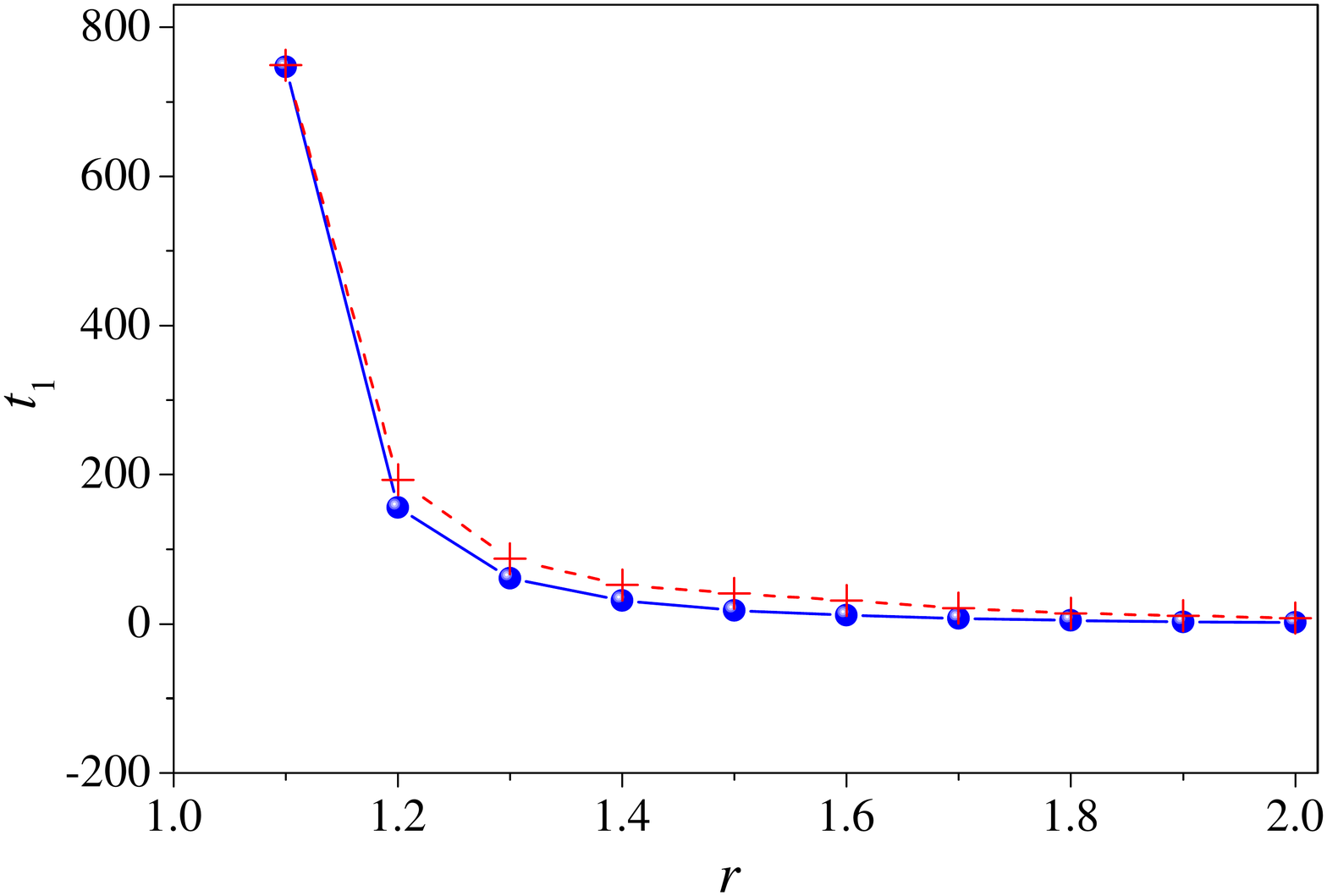}
    \caption{Comparison of the time scale $t_1$ in the 1D diatomic gas model that separates the kinetic and hydrodynamic processes obtained by Eq.~(9) analytically (red crisscrosses) and by simulations (blue bullets). }
    \end{figure}

Finally we present the results for variant 1D diatomic gas models. In Fig.~5(a)-(c), we compare the analytically predicted and simulated HCAFs for, respectively, the periodic model with repeating unit of $\mu_1-\mu_2-\mu_2$, the random diatom model with binary isotactic clusters of $\mu_1$ and $\mu_2$, and the random mass model with uniformly distributed masses $m_i\in (1,1.4)$. In all three cases, the analytical results fit the simulation results very well in the kinetic region $t<t_1$.
 
In summary, by introducing and tracing the tagged energy, we extend the kinetic approach to characterize the HCAF in the time region where the kinetic effect dominates. The system-size independent heat conductivity observed in previous studies, including its value and the system size range, is predicted quantitatively. Our study indicates that a full description of the HCAF should incorporate both the kinetic and hydrodynamic effects. Our method may be  applicable to other momentum conserving systems.

    \begin{figure*}
    \centering
    \includegraphics[width=18.2cm]{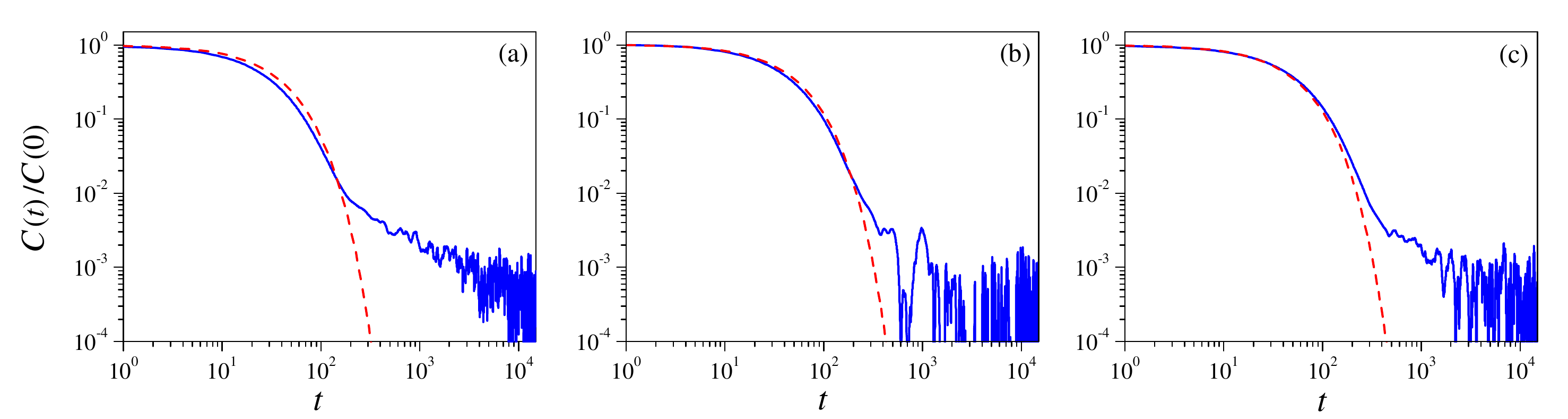}
    \caption{The heat current autocorrelation function for three variant 1D gas models. (a) Periodic diatom model with repeating $\mu_1-\mu_2-\mu_2$ unit. (b) Random diatomic gas model. In (a) and (b), $\mu_1=1.2$ and $\mu_2=1$. (c) Random gas model with the particle masses uniformly distributed between 1 and 1.4. In all the panels, the red dashed line is for the theoretical prediction given by Eq.~(6) and the black solid line is for the simulation result.}
    \end{figure*}

\bibliography{paper312}

\end{document}